\documentclass[useAMS,usenatbib]{mn2e}

 \usepackage{times}

\def\pcm3{{\rm\thinspace cm^{-3}}}

\def\contcaption{\@conttrue\SFB@caption\@captype}

\def\n_h{{\rm n_{H}}}

\def\NH1{{$N_{\rm HI}~$}}

\def\ga{{\rm\thinspace gauss}}

\def\approxlt{\mathrel{\hbox{\rlap{\lower .5ex \hbox {$\sim$}}
        \raise .15 ex \hbox{$<$}}}}
\def\approxgt{\mathrel{\hbox{\rlap{\lower .5ex \hbox {$\sim$}}
        \raise .15 ex \hbox{$>$}}}}

\def\la{\mathrel{\hbox{\rlap{\hbox{\lower4pt\hbox{$\sim$}}}\hbox{$<$}}}}
\def\ga{\mathrel{\hbox{\rlap{\hbox{\lower4pt\hbox{$\sim$}}}\hbox{$>$}}}}
\newbox\grsign \setbox\grsign=\hbox{$>$} \newdimen\grdimen
\grdimen=\ht\grsign
\newbox\simlessbox \newbox\simgreatbox \newbox\simpropbox
\setbox\simgreatbox=\hbox{\raise.5ex\hbox{$>$}\llap
     {\lower.5ex\hbox{$\sim$}}}\ht1=\grdimen\dp1=0pt
\setbox\simlessbox=\hbox{\raise.5ex\hbox{$<$}\llap
     {\lower.5ex\hbox{$\sim$}}}\ht2=\grdimen\dp2=0pt
\setbox\simpropbox=\hbox{\raise.5ex\hbox{$\propto$}\llap
     {\lower.5ex\hbox{$\sim$}}}\ht2=\grdimen\dp2=0pt
\def\simgreat{\mathrel{\copy\simgreatbox}}
\def\simless{\mathrel{\copy\simlessbox}}

\title[]{A massive white dwarf member of the Coma Berenices Open Cluster}
\author[P. D. Dobbie et al.]{P. D. Dobbie$^{1}$\thanks{E-mail:pdd@aao.gov.au},
S. L. Casewell$^{2}$, M. R. Burleigh$^{2}$ D.D. Boyce$^{2}$ \\ 
$^{1}$ Anglo-Australian Observatory, PO Box 296, Epping, NSW 1710, Australia \\
$^{2}$ Department of Physics and Astronomy, University of Leicester, University Road, Leicester LE1 7RH, UK\\ }

\begin{document}

\date{Accepted 2009 February 16. Received 2009 February 13; in original form 2008 December 31}

\pagerange{\pageref{firstpage}--\pageref{lastpage}} \pubyear{2002}

\maketitle

\label{firstpage}

\begin{abstract}

We report the identification, from a photometric, astrometric and spectroscopic study, of a massive white dwarf 
member of the nearby, approximately solar metalicity, Coma Berenices open star cluster (Melotte 111). We find the
optical to near-IR energy distribution of WD1216+260 to be entirely consistent with that of an isolated DA and 
determine the effective temperature and surface gravity of this object to be $T_{\rm eff}$=$15739^{+197}_{-196}$K and 
log~$g$=$8.46^{+0.03}_{-0.02}$. We set tight limits on the mass of a putative cool companion, M$\simgreat$0.036M$_{\odot}$ 
(spatially unresolved) and M$\simgreat$0.034M$_{\odot}$, (spatially resolved and a$\simless$2500AU). Based on the predictions
of CO core, thick-H layer evolutionary models we determine the mass and cooling time of WD1216+260 to be M$_{\rm WD}$=$0.90
\pm0.04$M$_{\odot}$ and $\tau$$_{\rm cool}$=$363^{+46}_{-41}$Myrs respectively. For an adopted cluster age of $\tau$=500$
\pm$100Myrs we infer the mass of its progenitor star to be M$_{\rm init}$=$4.77^{+5.37}_{-0.97}$M$_{\odot}$. We briefly discuss 
this result in the context of the form of the stellar initial mass-final mass relation.

\end{abstract}


\section{Introduction}

Galactic open star clusters have for decades been regarded as key entities with which to address important issues 
in stellar and galactic astrophysics e.g the form of the stellar initial mass function, the evolution of stellar 
angular momentum and the form of the initial mass-final mass relation. This is because fundamental stellar 
parameters such as age and composition can be more stringently constrained through studying a co-eval ensemble 
of objects at essentially the same distance (e.g. Meynet, Mermilliod \& Maeder 1993).

Melotte 111 (Coma Berenices cluster, RA=12$^{h}$22$^{m}$, DEC=+26$^{\circ}$, J2000.0) is the second closest open 
star cluster to the Sun. It appears to be marginally younger than the Hyades or Praesepe, with most recent age estimates 
lying in the range $\tau$=400-600Myrs (e.g. $\tau$$\approx$450Myrs, Bounatiro \& Arimoto 1992; $\tau$$\approx$500Myrs,
Odenkirchen et al. 2001; $\tau$$\approx$600Myrs, Kharchenko et al. 2005). Cayrel de Strobel (1990) has determined 
[Fe/H]=-0.065$\pm$0.021 from a high resolution spectroscopic study of a sample of eight F-,G- and K- type members. 
A similar but independent investigation of fourteen F and G dwarf associates has concluded that [Fe/H]=-0.052$\pm$0.047 
(Friel \& Boesgaard 1992). More recently, Gebran, Monier \& Richardo (2008), have determined [Fe/H]=0.07$\pm$0.09 through 
the high resolution spectroscopic study of eleven F dwarf members. Thus the cluster metalicity  is relatively close to the 
solar value. 

In the last decade there have been several attempts to determine the distance of Melotte 111. For example, using $Hipparcos$ parallax 
measurements, Van Leeuwen (1999) and Robichon et al. (1999) have estimated d=89.9$\pm$2.1pc and d=87.0$\pm$1.6pc, respectively. 
However, the reliability of both of these measurements has been challenged by Pinsonneault et al. (1998) and Makarov (2003). 
The former matched a theoretical isochrone to the cluster sequence in the $B$-$V$ colour to obtain d=80.9$\pm$1.5pc or d=84.1pc 
assuming [Fe/H]=-0.07$\pm$0.02 or solar metalicity respectively. The latter re-assessed the $Hipparcos$ intermediate 
astrometry data using a method which reduced the impact of errors in the along-scan attitude parameters to determine
d=80.6$\pm$1.6pc. However, from his new reduction of the raw $Hipparcos$ data, which included an improved treatment for the 
along-scan attitude issues and which is claimed to be superior to the Makarov (2003) approach, van Leeuwen (2007) has determined 
d=86.7$^{+1.0}_{-0.9}$pc. This is consistent with the Geneva photometry based estimate of Nicolet (1981), d=85.4$\pm$4.9pc and 
is closer to the Pinsonneault et al. (1998) determination, particularly if the cluster metalicity is nearer solar than originally 
assumed by these authors. As a consequence of this close proximity and the high galactic latitude of Melotte 111, levels of foreground 
reddening are very low e.g. E($B-V$)=0.006$\pm$0.013, Nicolet (1981); E($B-V$)$\le$0.0032, Taylor (2006). 

\begin{figure*}
\vspace{215pt}
\includegraphics{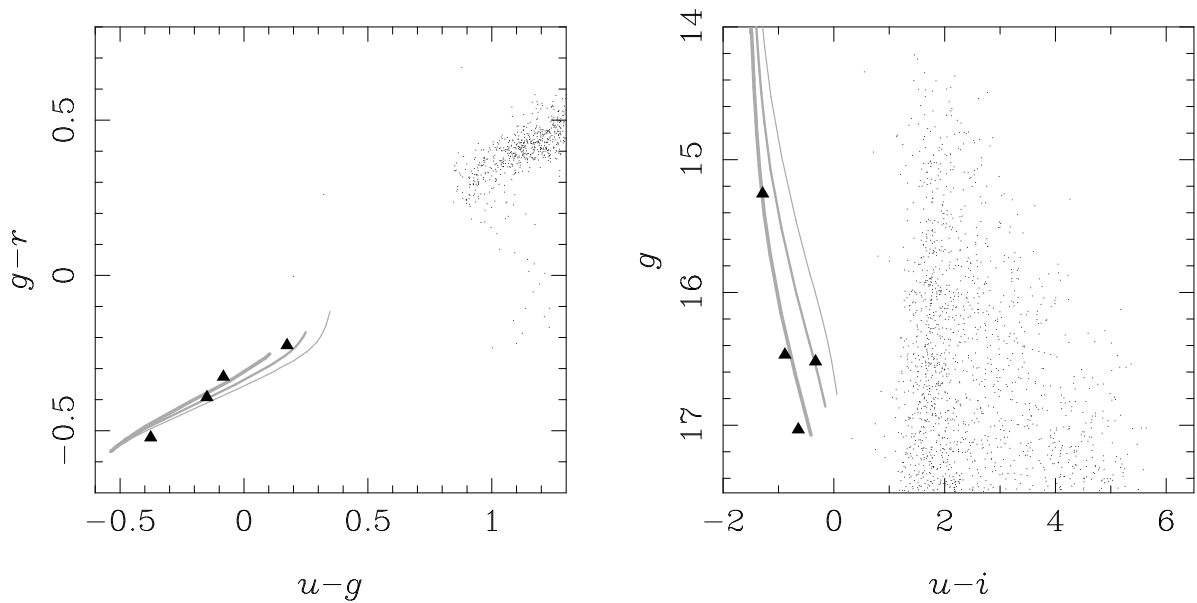}
\caption{Left: $u$$g$$r$ colour-colour diagram of stellar objects lying within the projected core radius of Melotte 111. Theoretical 
colour-colour tracks for DA white dwarfs with cooling times $\simless$600Myrs, from Holberg \& Bergeron (2006), are overplotted 
(thin grey line, 0.7M$_{\odot}$, medium grey line, 0.9M$_{\odot}$, thick grey line, 1.1M$_{\odot}$). Probable white dwarfs are highlighted 
(triangles). Right: $g$,$u$-$i$ colour-magnitude diagram  of the stellar objects lying within the projected core radius of Melotte 111. 
Theoretical white dwarf tracks are overplotted (as 
previously denoted). The locations in colour-magnitude space of the four probable white dwarfs (triangles) indicate that they are 
potentially members of Melotte 111.}
\end{figure*}

The mean heliocentric radial velocity of cluster members is close to r$_{\rm v}$$\approx$0kms$^{-1}$. For example, Abt \& Willmarth (1999) measured 
r$_{\rm v}$=-0.4$\pm$0.4 kms$^{-1}$ from sixteen stars. Kharchenko et al. (2005) determined r$_{\rm v}$=-1.17$\pm$0.99 kms$^{-1}$ 
by cross-referencing the ASCC-2.5 catalogue (Kharchenko et al. 2001) with the General Catalogue of Radial Velocities (Barbier-Brossat
\& Figon 2000). From CORAVEL spectroscopic observations of the G7 III + A2.5IV giant binary member Trumpler 91 (Trumpler 1938), Mermilliod,
Mayor \& Udry (2008) estimate r$_{\rm v}$=+1.06$\pm$1.5kms$^{-1}$. Despite being nearby, the tangential motion of members is rather 
small. From $Hipparcos$ data Robichon et al. (1999) determine $\mu_{\alpha}$cos $\delta$=$-$11.38$\pm$0.23mas yr$^{-1}$, $\mu_{\delta}$=$-
$9.05$\pm$0.12 mas yr$^{-1}$. This is corroborated by the location of the clump of stars brighter than $J$=11 seen in a UCAC2 proper motion 
vector point diagram for the region of sky centered on the cluster (Figure 1 of Casewell et al. 2006). Kraus \& Hillenbrand (2007) have 
recently performed a detailed study of the structure and the mass of Melotte 111 utilising both astrometry and photometry. They have concluded 
that the projected tidal radius is 4.3$\pm$0.2$^{\circ}$, which translates to 6.4$\pm$0.3pc at a distance of d=85pc, and that the total 
cluster mass is M$_{\rm tot}$=112$\pm$16M$_{\odot}$.

To date, there have been very few efforts to identify and study the white dwarf population of Melotte 111. This is presumably because the 
cluster is relatively sparse and its members do not have a distinctive proper motion. From a $UBV$ photographic survey of 21.2 sq. degrees 
of sky centered on the cluster down to m$_{\rm pg}$$\approx$17, Stephenson (1960) identified five candidate white dwarf members, but none of 
these appears to have yet been subject to detailed follow-up study. Brosch et al. (1998) used the $FAUST$ ultraviolet imaging telescope to 
observe a 67.5 sq. degree region of sky centered on the Coma galaxy cluster (RA=12$^{h}$58$^{m}$, DEC=+28$^{\circ}$05$^{m}$, J2000.0) and 
claims to have unearthed $\sim$5-10 candidate white dwarf members of Melotte 111. However, these objects all lie well beyond the tidal radius
of the cluster and thus are clearly not gravitationally bound members.  

Nevertheless, based on scaling the number of known white dwarf members of the slightly older, more massive Hyades (M$_{\rm tot}$=300-460M$
_{\odot}$, Pels et al. 1975, Reid  1992; N$_{\rm WD}$$\approx$10, von Hippel 1998) and Praesepe (M$_{\rm tot}$=500-600M$_{\odot}$,
Adams et al. 2002, Kraus \& Hillenbrand 2007; N$_{\rm WD}$$\approx$10, Casewell et al. 2008) clusters, we should expect a photometric 
survey of Melotte 111 to unearth $\sim$2 degenerate members. The detailed investigation of the white dwarf members of open star clusters
(e.g. Sweeney 1976, Weidemann 1977, Weidemann 2000, Williams et al. 2004, Kalirai et al. 2005, Dobbie et al. 2006a) is arguably the best 
observational approach to ascertaining the form of the stellar initial mass-final mass relation (IFMR). The difference between the age of an
open star cluster and the cooling time of a white dwarf member provides an estimate of the lifetime of the progenitor star. The mass of the 
progenitor star can then be estimated by comparing this lifetime to the predictions of stellar evolutionary calculations. Robust constraints 
on the form of the IFMR are very useful to several threads of astrophysical research. For example, the relation is an important component of 
models of the chemical evolution of the Galaxy as it provides an estimate of the amount of gas, enriched with C, N and other metals, a low or
intermediate mass star returns to the ISM (e.g. Carigi, Colin \& Peimbert 1999). Understanding the form of the IFMR is crucial to
deciphering information locked up in the white dwarf luminosity functions of stellar populations (e.g. Jeffery et al. 2007, Oswalt et al. 1996).
Furthermore, the shape of the upper end of the IFMR is of particular interest as it places limits on the minimum mass of star that will experience
a core-collapse supernova explosion. With a tight handle on this mass, for example, the observed diffuse SNe neutrino background 
can serve as an empirical normalisation check on estimates of the star formation history of the universe (e.g. Hopkins et al. 2006).

\begin{table*}
\begin{minipage}{162mm}
\begin{center}
\caption{A summary of the astrometric and photometric properties of the four candidate white dwarf members of Melotte 111 unearthed by our SDSS 
based imaging survey of the central cluster region.}
\label{wdmass}
\begin{tabular}{lccccccc}
\hline
\multicolumn{1}{c}{Object} & Stephenson & $u$ & $g$ & $r$ & $i$ &  $\mu_{\alpha}$cos $\delta$ & $\mu_{\delta}$ \\ 

   & ID No.  & \multicolumn{4}{c}{SDSS} & (mas yr$^{-1}$) & (mas yr$^{-1}$) \\
\hline

SDSS J121635.01+262934.7 & 1 & 14.88$\pm$0.03 & 15.26$\pm$0.02 & 15.78$\pm$0.02 & 16.16$\pm$0.01 & -29.0$\pm$10.3 & -13.2$\pm$9.2 \\
SDSS J121845.69+264831.7 & 2 & 16.32$\pm$0.01 & 16.47$\pm$0.01 & 16.86$\pm$0.02 & 17.20$\pm$0.03 & 61.2$\pm$58.8 & -55.4$\pm$57.0 \\
SDSS J121856.17+254557.1 & 3 & 16.69$\pm$0.01 & 16.52$\pm$0.02 & 16.75$\pm$0.01 & 17.03$\pm$0.02 & -9.7$\pm$7.9 & -8.2$\pm$8.2 \\
SDSS J122420.51+264738.9 & - & 16.95$\pm$0.03 & 17.03$\pm$0.02 & 17.36$\pm$0.02 & 17.59$\pm$0.03 & 2.5$\pm$6.2 & -10.0$\pm$6.1 \\

\hline
\end{tabular}
\end{center}
\end{minipage}
\end{table*}

Driven by a desire to improve current observational constraints on the form of the IFMR and the availability of high quality multiband 
photometry from the Sloan Digitial Sky Survey (SDSS) for Melotte 111, we have undertaken a search of the central regions of the cluster for
white dwarf members. In the next section we outline our photometric and astrometric survey of the cluster center. We then describe the acquisition
and analysis of our spectroscopic follow-up data and the use of this to assess the membership status of candidates. Finally, we examine a likely 
white dwarf member in the context of the IFMR.

\section{A search for potential white dwarf members of Melotte 111}

\subsection[]{Optical photometry}

Melotte 111 lies entirely within the footprint of the SDSS data release 6 (DR6), so for this region of sky we have at our disposal 
optical photometry in five bands, $u$$g$$r$$i$$z$, with a precision of up to 2-3\% (Adelman-McCarthy et al. 2008). Harris et al. (2003)
have demonstrated that single white dwarfs with $T_{\rm eff}$$\simgreat$12000K are clearly separated from main sequence objects and 
quasars in a $u$$g$$r$ colour-colour diagram. Therefore, as a first step, we have extracted from the SDSS DR6 Data Archive Server, clean 
$u$,$g$ and $r$ band photometry for objects classified as stellar lying within 1.5$^{\circ}$ of the cluster center. This corresponds 
roughly to the projected $e$-folding radius for the surface density of 0.5-1.0M$_{\odot}$ members (Kraus \& Hillenbrand 2007). We have imposed 
a brightness criterion of $g$$\le$17.5 since given the age and the distance of the cluster we would not expect to find white dwarf members 
below this magnitude limit (e.g. Bergeron et al. 1995). The  $u$$g$$r$ colour-colour diagram for these data is shown in Figure 1, left. 
Model colours (Bergeron et al. 1995, Holberg \& Bergeron 2006) for 0.7, 0.9 and 1.1M$_{\odot}$ DA white dwarfs, with $\tau$$_{\rm cool}$$\le
$600Myrs, are overplotted (thin grey, medium grey and thick grey line respectively). Inspection of this plot reveals that four objects 
clearly stand out as candidate white dwarfs (filled triangles). Nevertheless,  from this analysis alone we cannot determine if these are cluster 
members since field white dwarfs and some hot subdwarfs can also reside in this part of colour-colour space. In an attempt to exclude 
field contaminants we have examined their location in the $u$-$i$,$g$ colour-magnitude diagram (Figure 1, right). Once again we have 
overplotted synthetic photometric tracks (Bergeron et al. 1995, Holberg \& Bergeron 2006) for 0.7, 0.9 and 1.1M$_{\odot}$ DA white dwarfs 
with $\tau$$_{\rm cool}$$\le$600Myrs, here placed at d=85pc (denoted as in the colour-colour plot). However, given the effects of the finite 
depth of Melotte 111 (r$_{\rm tidal}$=6.4$\pm$0.3pc) and the proximity of the four objects to these model tracks, at this stage we are led to 
retain all as candidate white dwarf cluster members. We note that our three brightest objects were previously identified as potential members of 
Melotte 111 by Stephenson (1960). The coordinates and the $u$$g$$r$$i$ photometry for all four candidates can be found in Table 1. 

\begin{figure}
\vspace{200pt}
\includegraphics{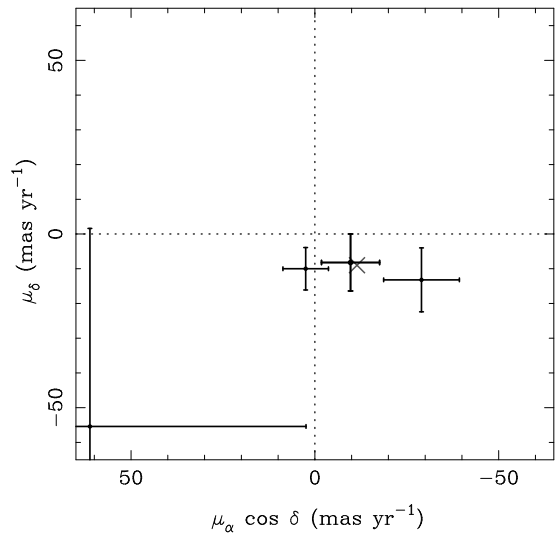}
\caption{A proper motion vector point diagram, based on SuperCOSMOS data obtained from the Survey Science Archive, for the photometrically 
selected candidate white dwarf members of Melotte 111. The mean proper motion of cluster members, as determined by Robichon et al. 
(1999) using Hipparcos data, is highlighted ('$\times$'). As the proper motion of SDSS J122420.51+264738.9 lies $\simgreat$2.2$\sigma$ from the 
cluster mean we conclude that it is likely to be a field star. The proper motions of the remaining three potential white dwarfs can be considered 
at least marginally consistent with Melotte 111.}
\end{figure}

\begin{figure*}
\vspace{195pt}
\includegraphics{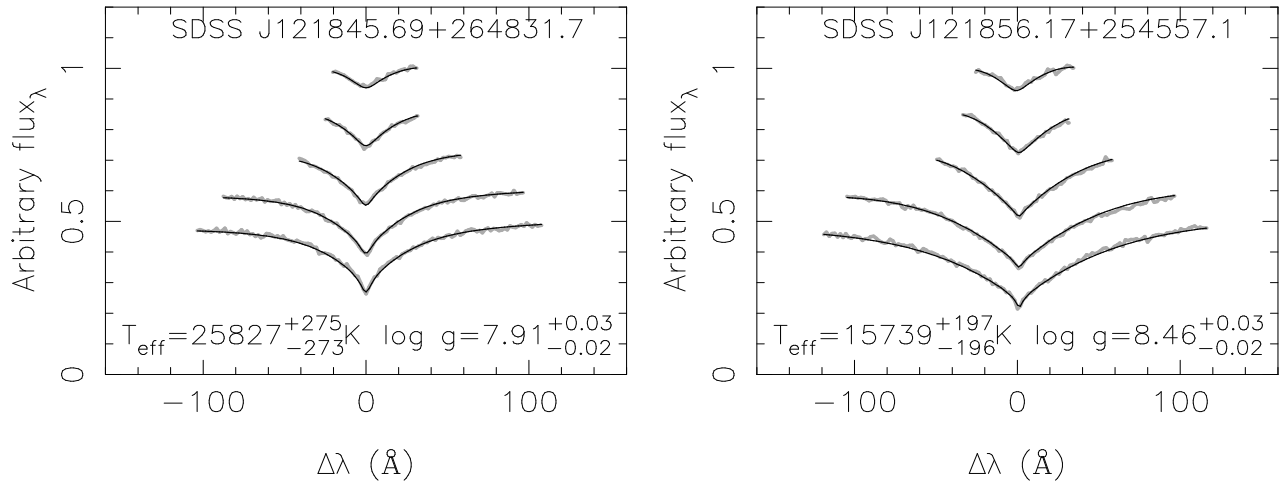}
\caption{The results of fitting synthetic profiles (thin black lines) to the observed Balmer lines, H-$\beta$ to H-8 (thick grey lines). 
The flux$_{\lambda}$ units are arbitrary.}
\end{figure*}

\subsection[]{Proper motions} 

As a further aid to discriminating between cluster members and the field population, we have extracted proper motions for the four candidate white 
dwarfs from the SuperCOSMOS Science Archive (Table 1). These measurements, which are plotted in Figure 2, have been determined with respect to background 
galaxies and, like the Hipparcos astrometry (e.g. Robichon et al. 1999), are effectively absolute (see Hambly et al. 2001 for further details). Although 
the small proper motion of the cluster dictates that these SuperCOSMOS data are not of sufficient accuracy to compellingly demonstrate that a particular 
candidate is an associate of Melotte 111, an object must have astrometric properties at least consistent with the cluster to be considered a member. Indeed,
since SDSS J122420.51+264738.9 has a proper motion lying $\simgreat$2.2$\sigma$ from the Hipparcos determination of the cluster mean ('$\times$' in Figure 2)
we conclude that it is likely to be a field star and consequently do not consider it further in our analysis. While SDSS J121856.17+254557.1 is the only 
candidate with a proper motion within 1$\sigma$ of that of Melotte 111, the measurements for SDSS J121635.01+262934.7 and SDSS J121845.69+264831.7 are 
at least marginally consistent with that of the cluster and do not rule out the possibility that either is a member.
However, a recent spectroscopic analysis of SDSS J121635.01+262934.7 (PG1214+268) has confirmed it to be a DA white dwarf with M$\approx$0.6M$_{\odot}$ and 
$T_{\rm eff}$$\approx$65000K, residing at d$\sim$315pc (Liebert, Bergeron \& Holberg 2005). Thus we can say with confidence that it is not a member of Melotte 
111 and exclude it from further discussion. Clearly, spectroscopic data is required to examine more thoroughly the nature and the membership status of our 
two remaining candidates.

\section{Spectroscopic follow-up of SDSS J121845.69+264831.7 and SDSS J121856.17+254557.1}

\subsection[]{The WHT observations}

We have obtained optical spectroscopy of both SDSS J121845.69+264831.7 and SDSS J121856.17+254557.1 using the William Herschel Telescope (WHT) and the double-armed
Intermediate dispersion Spectrograph and Imaging System (ISIS). These observations were conducted on 2006/02/02 and 2008/07/25, when excellent conditions prevailed. 
On both nights the sky was very clear with seeing $\sim$0.6-0.9'', while additionally on 2008/07/25, humidity levels were very low ($\sim$10\%). For the observation 
of SDSS J121845.69+264831.7, ISIS was configured with a 1.0'' slit and data covering $\lambda$$\approx$3600-5500\AA\ were acquired using the R300B grating on the blue 
arm only ($\lambda$/$\delta\lambda$$\approx$1200). For the observation of SDSS J121856.17+254557.1, the spectrograph was configured with a 0.6'' slit and data covering
the two wavelength ranges, $\lambda$$\approx$3600-5500\AA, 6200-7000\AA, were obtained simultaneously using the R300B ($\lambda$/$\delta\lambda$$\approx$2000) and R1200R
($\lambda$/$\delta\lambda$$\approx$10000) gratings on the blue and red arms respectively. We performed a series of short integrations to obtain total exposure times of
20 and 45 minutes on SDSS J121845.69+264831.7 and SDSS J121856.17+254557.1 respectively. The CCD frames were debiased and flat fielded using the IRAF procedure CCDPROC.
Cosmic ray hits were removed using the routine LACOS SPEC (van Dokkum 2001). Subsequently, the spectra were extracted using the APEXTRACT package and wavelength calibrated 
by comparison with the CuAr+CuNe arc spectra taken immediately before and/or after the science exposures. The removal of remaining intrument signature from the science 
spectra obtained on the first night was undertaken using an observed and a synthetic spectrum (Bohlin 2000) of G191-B2B. Data obtained on the second night were corrected 
using an observation of the bright DC white dwarf WD1918+386. The reduced spectra clearly show both objects to be hydrogen rich white dwarfs, ie. WD1216+270 and WD1216+260.

\subsection[]{The model atmosphere calculations}

We have used recent versions of the plane-parallel, hydrostatic, non-local thermodynamic equilibrium (non-LTE)
atmosphere and spectral synthesis codes TLUSTY (v200; Hubeny 1988, Hubeny \& Lanz 1995) and SYNSPEC (v48; Hubeny, 
I. and Lanz, T. 2001, http://nova.astro.umd.edu/) to generate a grid of pure-H synthetic spectra covering the 
effective temperature and surface gravity ranges $T_{\rm eff}$=15000-30000K and log~$g$=7.5-8.75 respectively. We have 
employed a model H atom incorporating the 8 lowest energy levels and one superlevel extending from n=9 to n=80, where the 
dissolution of the high lying levels was treated by means of the occupation probability formalism of Hummer \& Mihalas (1988),
generalised to the non-LTE situation by Hubeny, Hummer \& Lanz (1994). The calculations assumed radiative equilibrium (since 
our test calculations and those of the referee, Detlev Koester, indicate that in this effective temperature regime convection 
transports $\simless$1\% of the flux and its neglect has a neglible impact on the emergent spectrum) and included the bound-free
and free-free opacities of the H$^{-}$ ion and incorporated a full treatment for the blanketing effects of HI lines and the Lyman
$-\alpha$, $-\beta$ and $-\gamma$ satellite opacities as computed by N. Allard (e.g. Allard et al. 2004). During the calculation 
of the model structure the hydrogen line broadening was addressed in the following manner: the broadening by heavy perturbers 
(protons and hydrogen atoms) and electrons was treated using Allard's data (including the quasi-molecular opacity) and an 
approximate Stark profile (Hubeny, Hummer \& Lanz 1994) respectively. In the spectral synthesis step detailed profiles for the 
Balmer lines were calculated from the Stark broadening tables of Lemke (1997).

\begin{table*}
\begin{minipage}{155mm}
\begin{center}
\caption{Alternative designations, effective temperatures, surface gravities, predicted absolute $g$ magnitudes and distance modulii
for SDSS J121845.69+264831.7 and SDSS J121856.17+254557.1.}
\label{wdmass}
\begin{tabular}{cclcrc}
\hline
Object & Other IDs & $T_{\rm eff}$$^{*}$ & log~$g$$^{*}$ &  M$_{g}$ & (m-M)$_{g}$ \\ 
\hline
\multicolumn{1}{l}{SDSS J121845.69+264831.7} &  \multicolumn{1}{l}{LB4, WD1216+270}                   &  $25827^{+275}_{-273}$ & $7.91^{+0.03}_{-0.02}$ & $9.94^{+0.11}_{-0.12}$ & $6.53^{+0.12}_{-0.11}$  \\   
\multicolumn{1}{l}{SDSS J121856.17+254557.1} &  \multicolumn{1}{l}{Ton 607, CSI+26-12165, WD1216+260} &  $15739^{+197}_{-196}$ & $8.46^{+0.03}_{-0.02}$ & $11.84^{+0.12}_{-0.12}$ & $4.68^{+0.12}_{-0.12}$  \\

\hline
\end{tabular}
\end{center}
$^{*}$ Formal fit errors. 
\end{minipage}
\end{table*}

\subsection[]{Determination of the effective temperature and surface gravity}

As in our previous work (e.g. Dobbie et al. 2006a), comparison between the models and the low resolution data is 
undertaken using the spectral fitting program XSPEC (Shafer et al. 1991). In our analysis all lines from H-$\beta$ to
H-8 are included in the fitting process. XSPEC works by folding a model through the instrument response before comparing 
the result to the data by means of a $\chi^{2}-$statistic. The best fit model representation of the data is found by 
incrementing free grid parameters in small steps, linearly interpolating between points in the grid, until the value of $\chi^{2}$ 
is minimised. Errors in the $T_{\rm eff}$s and log~$g$ s are calculated  by stepping the parameter in question away from 
its optimum value and re-determining minimum $\chi^{2}$ until the difference between this and the true minimum $\chi^{2}$ 
corresponds to $1\sigma$ for a given number of free model parameters (e.g. Lampton et al. 1976). The results of our 
fitting procedure are given in Table 2 and shown overplotted on the data in Figure 3.  It should be noted that the 
parameter errors quoted here are formal $1\sigma$ fit errors and undoubtedly underestimate the true uncertainties. For our 
subsequent calculations we adopt more realistic levels of uncertainty of 2.3\% and 0.07dex in effective temperature and 
surface gravity respectively (e.g. Napiwotzki, Green \& Saffer 1999).

\subsection{Distance, radial velocity and cluster membership status}

Using the effective temperature and the surface gravity estimates listed in Table 2, and the model white dwarf photometry of Bergeron 
et al. (1995), which has been updated by Holberg \& Bergeron (2006), we have derived the absolute $g$ magnitudes (M$_{g}$) of WD1216+270 
(SDSS J121845.69+264831.7) and WD1216+260 (SDSS J121856.17+254557.1). Subsequently, we have determined the distance modulii of these two objects,
neglecting extinction which is minimal along this line of sight. These are plotted in Figure 4 (solid bars), along with reasonable estimates for 
the distance and the tidal bounds of Melotte 111 (d=85pc, r$_{\rm tidal}$=6.4pc). Inspection of this plot reveals that while WD1216+270 is probably 
a field star residing some way behind the cluster, the distance modulus of WD1216+260 argues rather strongly that it is associated with Melotte 111.

\begin{figure}
\vspace{97pt}
\includegraphics{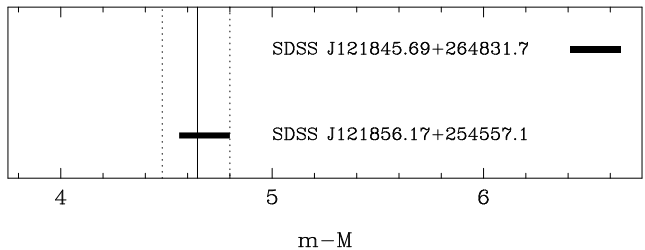}
\caption{The estimated distance modulii of WD1216+270 (SDSS J121845.69+264831.7) and WD1216+260 (SDSS J121856.17+254557.1) based on the observed SDSS 
photometry and the synthetic photometry from Holberg \& Bergeron (2006). The distance of the centre (d=85pc) and the tidal limits (r$_{\rm tidal}$=6.4pc) of the 
cluster are overplotted (solid and dotted vertical lines respectively). WD1216+260 resides within the tidal bounds of Melotte 111.}
\end{figure}

\begin{figure}
\vspace{180pt}
\includegraphics{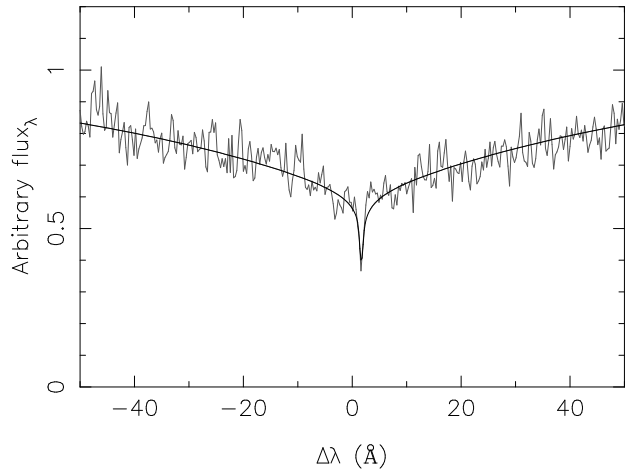}
\caption{The results of our fitting of a synthetic profile (thin black line) to the central portions of the observed H-$\alpha$ Balmer line of 
WD1216+260 (thick grey lines). The flux$_{\lambda}$ units are arbitrary.}
\end{figure}

To further scrutinise the membership status of this one remaining candidate (WD1216+260), we have acquired a radial velocity measurement. We have removed
the effects of telluric water vapour from the red arm ISIS spectroscopic data using a template absorption spectrum. Custom written IDL routines have been used to 
normalise both the observed H-$\alpha$ line and the profile from a synthetic spectrum corresponding to $T_{\rm eff}$=15750K and log~$g$=8.45. Subsequently, 
the model has then been compared to the data, where the Levenberg-Marquardt algorithm has been used to minimise a $\chi$$^{2}$ fit statistic. The result 
of this process is displayed in Figure 5. The line velocity shift, after correction to the heliocentric rest frame, is shown in Table 3. The uncertainty
in this measurement has been estimated using the bootstrapping method of statistical resampling (Efron 1982). The gravitational 
redshift component of the velocity shift was then determined using Equation 1, where $M$ and $R$ are the mass and radius of the white dwarf in solar units 
respectively and $v$ is the gravitational redshift in kms$^{-1}$.

\begin{equation}
v = 0.635 M / R
\end{equation}

The mass and radius have been determined using modern evolutionary tracks supplied by the Montreal group (e.g. Fontaine, Brassard \& Bergeron 2001),
more specifically the calculations which assume a mixed CO core and thick H surface layer structure. The radial velocity is represented by the 
difference between the measured shift of the line and the calculated gravitational redshift. It can be seen from Table 3 that the radial velocity of 
WD1216+260 is very close to the value expected for a member of Melotte 111 and entirely consistent within the measurement uncertainties. On the basis 
of the projected location, the proper motion, the distance and the radial velocity of WD1216+260, we conclude that it is a white dwarf member of Melotte 111.

\section{A white dwarf member of Melotte 111}

\begin{table*}
\begin{minipage}{168mm}
\begin{center}
\caption{Additional physical details of WD1216+260 (SDSS J121856.17+254557.1), the white dwarf probable member of Melotte 111. The mass and the cooling time 
have been estimated using the mixed CO core composition ``thick H-layer'' evolutionary calculations of the Montreal Group (e.g. Fontaine, Brassard 
\& Bergeron 2001).}

\label{wdmass}
\begin{tabular}{ccccccccccc}
\hline
Object & M$_{\rm WD}$ &  R$_{\rm WD}$ & rv$_{\rm Mel 111}$ & H-$\alpha$ shift & GR & rv$_{\rm WD}$ & $\tau_{c}$ & M$_{\rm init}$ \\ 
& \multicolumn{1}{c}{(M$_{\odot}$)} & \multicolumn{1}{c}{(R$_{\odot}$$\times$10$^{-3}$)} & \multicolumn{4}{c}{(kms$^{-1}$)} & \multicolumn{1}{c}{(Myrs)} & \multicolumn{1}{c}{(M$_{\odot}$)} \\ \hline

\multicolumn{1}{l}{SDSS J121856.17+254557.1} & $0.902\pm0.045$ & 9.26$\pm$0.84 &  -0.4$\pm$0.4 & 57.1$^{+0.5}_{-2.5}$ & 61.9$\pm$6.4 &  -4.8$^{+6.4}_{-6.9}$  & $363^{+46}_{-41}$ &  $4.77^{+5.37}_{-0.97}$\\
\hline
\end{tabular}
\end{center}
\end{minipage}
\end{table*}

\subsection{The near-IR energy distribution of WD1216+260}

We note that due to its proximity and intrinsic luminosity WD1216+260 was detected in the 2MASS All Sky Survey (Skrutskie et al. 2006), albeit weakly and only in the shortest waveband, $J$. It is listed in the Point Source Catalogue as having 
$J_{\rm 2MASS}$=16.79$\pm$0.17. Furthermore, it was detected at $K$  in the course of the Galactic Clusters Survey (GCS) component of the United Kingdom Infrared 
Digital Sky Survey (UKIDSS; Lawrence et al. 2007). Data extracted from the Wide Field Camera Science Archive (WSA; Hambly et al. 2008) indicates that for this 
object, $K_{\rm WFCAM}$=17.08$\pm$0.09. UKIDSS $J$ and $H$ band photometry are not yet available for the Melotte 111 region of sky.

\begin{figure}
\vspace{185pt}
\includegraphics{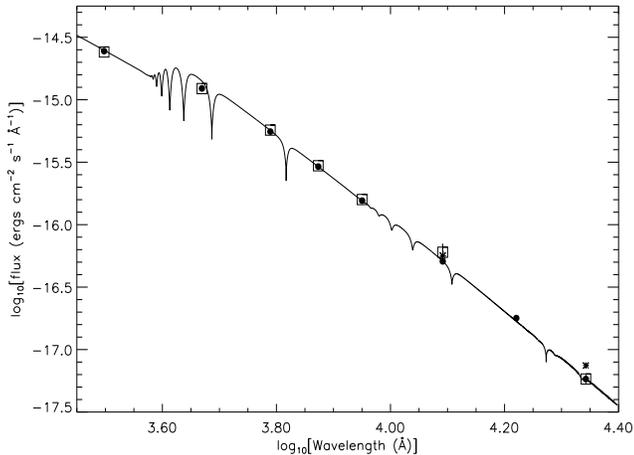}
\caption{The observed optical (SDSS $u$,$g$,$r$,$i$ and $z$) to near-IR (2MASS PSC $J$ and UKIDSS GCS $K$) spectral energy distribution of WD1216+260 (open squares and error bars). The model photometry of Bergeron et al. (1995) as updated by Holberg \& Bergeron (2006) is overplotted (solid circles), where we have assumed $T_{\rm eff}$=15739K and log~$g$ = 8.46 and a distance modulus of (m-M)$_{g}$=4.68. A TLUSTY/SYNSPEC synthetic spectrum for $T_{\rm eff}$=15739K and log~$g$ = 8.46 and scaled to match the observed optical photometry, is also overplotted (solid line). The estimated $J$ and $K$ flux levels resulting from the presense of a putative spatially unresolved M=0.036M$_{\odot}$ brown dwarf companion are shown (asterisks).}
\end{figure}

We have used data listed in Holberg \& Bergeron (2006; SDSS $u$$g$$r$$i$$z$), Cohen et al (2003; 2MASS $J$) and Hewett et al (2006; WFCAM $K$) to convert 
the optical and near-IR magnitudes to fluxes and have plotted these in Figure 6 (open squares). Model DA white dwarf photometry, from Bergeron et al. (1995) as updated 
by Holberg \& Bergeron (2006), is overplotted (solid circles), where we have assumed $T_{\rm eff}$=15739K and log~$g$ = 8.46 and a distance modulus of (m-M)$_{g}$=4.68.
Note that the model $K$ magnitude has been moved from the 2MASS system to the Wide Field Camera (WFCAM) system using transform equations from Hodgkin et al. (2009), here 
extrapolated slightly to the blue. Additionally we have overplotted a TLUSTY/SYNSPEC synthetic spectrum of the appropriate effective temperature and surface gravity, which 
has been scaled to match the observed optical fluxes (solid line).

Approximately 25\% of DAs in the effective temperature regime occupied by WD1216+260 are metal-rich (ie. DAZs; Zuckerman et al. 2003) and at least 14\% of 
these are observed, through the existence of an infrared excess (e.g. Kilic et al. 2005, Becklin et al. 2005), to harbour a dust disk which may be 
associated with a former planetary system (Kilic \& Redfield 2007; DAZd). The discovery of a DAZd in an open cluster, where the population age and progenitor 
mass can be relatively well constrained, would be of particular interest. However, the good agreement between the observed and the theoretical infrared fluxes means that the current data 
provide no evidence for the existence of a dust disk around WD1216+260. Moreover, these data do not indicate the presense of a close (spatially unresolved) cool low mass 
companion (e.g. Farihi \& Christopher 2004, Dobbie et al. 2005, Maxted et al. 2006, Burleigh et al. 2006).

Since WD1216+260 is a comparatively massive object with a likely total age of $\tau$$\sim$500Myrs, we are able to set a fairly tight upper limit on the mass of a putative 
cool companion. To do this we first moved the mean CIT magnitudes for L and T dwarfs listed in Table 9 of Vrba et al. (2004) onto the 2MASS ($J$) and Mauna Kea Observatories 
($K$; MKO) sytems using published fits relating spectral type and magnitude offsets between photometric systems (Stephens \& Leggett 2004). We assumed here that $K$$_{\rm 
MKO}$$\approx$$K$$_{\rm WFCAM}$, since WFCAM employs the MKO filter set (e.g. Hewett et al. 2006) and Leggett et al. (2006) find no compelling evidence for a colour term at 
$K$ between the UFTI MKO system (e.g. Hawarden et al. 2001) and the WFCAM system (although their data is rather limited). We then constructed the relations, spectral 
type versus luminosity, effective temperature and near-IR magnitudes using the Vrba et al. (2004) data, where the COND03 models of Baraffe et al. (2003) were used to estimate
the upwards offset in luminosity between this field dwarf dominated sample and the younger open cluster population. As we were aiming to set an upper limit on the mass of a
putative companion, we assumed $\tau$=600Myrs for Melotte 111 (the upper bound of the age estimate) and further offset the predicted magnitudes by +0.4 to account for the 
dispersion observed in the near-IR photometry of L/T dwarfs of a given spectral type (e.g. Tinney et al. 2003). Subsequently, beginning with T8 we combined progressively 
earlier and earlier spectral types with the model white dwarf photometry until the predicted flux at $K$ matched the 3$\sigma$ upper limit on the observed flux ($K_{\rm
WFCAM}$=16.81). This occured at spectral type T2.5 ($T_{\rm eff}$$\sim$1400K), which for our adopted age corresponds to M=0.036M$_{\odot}$.

Additionally, we inspected a 1'$\times$1' (representing a projected area of 5000AU$\times$5000AU) WFCAM $K$ band thumbnail image centered on WD1216+260. We found that this white 
dwarf is the only significant detection within this region. Photometry for other catalogued point sources within 30' indicate that a $\ge$3$\sigma$ detection corresponds to $K$$
\simless$18.65. At the distance of Melotte 111 and $\tau$=600Myrs this is consistent with M$\simgreat$0.034M$_{\odot}$. Thus we conclude that WD1216+260 has no spatially resolved 
cool companion with M$\simgreat$0.034M$_{\odot}$ within a$\sim$2500AU and no spatially unresolved cool companion with M$\simgreat$0.036M$_{\odot}$. Farihi, Becklin \& Zuckermann (2005)
estimated that $\simless$0.5\% of white dwarfs have brown dwarf companions with M$\simgreat$0.02-0.05M$_{\odot}$ within a$\simless$5000AU. Thus, this result is not entirely 
unexpected.

\subsection{In the context of the IFMR}

As a member of an open star cluster and having probably evolved essentially as a single star, WD1216+260 is potentially useful for constraining the form
of the IFMR. Therefore, we have used the Montreal group evolutionary tracks to determine a cooling time, where cubic splines have 
been used to interpolate between points in this grid. The results of this process are shown in Table 3. The lifetime of the progenitor star has been derived
by subtracting this cooling time from the adopted cluster age, where, for the reasons outlined in the Section 1, it has been assumed that $\tau$=500$\pm$100Myrs.
Subsequently, we have 
used cubic splines to interpolate between the lifetimes calculated for stars of solar composition by Girardi et al. (2000) and have constrained the 
progenitor mass to the value shown in the final column of Table 3. The error we quote in the progenitor mass takes into account the spread in recent cluster 
age determinations and the uncertainty in the cooling time of the white dwarf, but should be regarded as an illustrative rather than a robust estimate.

We find the location of WD1216+260 in initial mass-final mass space to be consistent with the general trend outlined by the bulk of white dwarf members of 
near solar metalicity (ie. -0.15$\simless$[Fe/H]$\simless$+0.15) open clusters (e.g. 
Kalirai et al. 2008 - NGC6819 and NGC7789,  Claver et al. 2001 - Hyades and Praesepe, Casewell et al. 2009 - Praesepe, Kalirai et al. 2005 - NGC2099, 
Williams \& Bolte 2007 - NGC6633, Rubin et al. 2008 - NGC1039, Dobbie et al. 2009 - NGC2287 and NGC3532, Koester \& Reimers 1996 - NGC2516 and Dobbie et
al. 2006a,b - Pleiades). While the mass range for the progenitor star of WD1216+260 is broad, 
primarily due to the substantial uncertainty on the age of Melotte 111 (e.g. Salaris et al. 2008), the most probable initial and final masses are very similar to 
those of the two white dwarf members of NGC2287 and the most massive degenerate in NGC1039. Therefore, WD1216+260 lends some support, albeit rather weak 
given the large uncertainties, to our conclusion in Dobbie et al. (2009) that the cluster data appear indicate that the IFMR is less steep at M$_{\rm init}$
$>$4M$_{\odot}$ than in the range 3M$_{\odot}$$\simless$M$_{\rm init}$$\simless$4M$_{\odot}$. As discussed at length in Dobbie et al. (2009), this decrease 
in the gradient of the IFMR is expected on theoretical grounds. The form of the initial mass versus the core mass at the time of the first thermal pulse 
relation (Karakas, Lattanzio \& Pols 2002) and a recent theoretical IFMR (Marigo \& Girardi 2007), reasonably reproduce that of the data at M$_{\rm init}$$>$3M$
_{\odot}$, although admitedly with offsets in final mass. Additional and/or better data in the initial mass regime M$_{\rm init}$$\simgreat$5M$_{\odot}$ would help 
to consolidate this conclusion.

\section{Summary}

We have utilised SDSS photometry, SuperCOSMOS proper motions and optical spectroscopy to search for white dwarf members within the 
central regions of the open cluster Melotte 111. We have identified one probable degenerate member, WD1216+260, with a mass of M$
_{\rm WD}$=$0.90\pm0.04$M$_{\odot}$. The optical to near-IR energy distribution of this object, as described by SDSS, 2MASS and UKIDSS GCS data,
is entirely consistent with that of an isolated DA white dwarf. We have set tight upper limits on the mass of a putative low mass companion, 
M$\simgreat$0.036M$_{\odot}$ (spatially unresolved) and M$\simgreat$0.034M$_{\odot}$, (spatially resolved and a$\simless$2500AU).
Moreover, for an adopted cluster age of $\tau$=500$\pm$100Myrs, we have infered the mass of its progenitor star to be M$_{\rm init}$=$4.77^
{+5.37}_{-0.97}$M$_{\odot}$. The location of WD1216+260 in initial mass-final mass space is consistent with a near-monotonic positive correlation
outlined by the bulk of open cluster white dwarfs and our earlier conclusion that the these objects appear to indicate that the IFMR is less 
steep at M$_{\rm init}$$>$4M$_{\odot}$ than in the range 3M$_{\odot}$$\simless$M$_{\rm init}$$\simless$4M$_{\odot}$, at least for metalicity close
to the solar value.

\section*{Acknowledgments}

SLC and DB are funded by STFC grants. MBU is supported by a STFC advanced fellowship. We thank Dan Bramich and Neil O'Mahony for supporting the 
WHT observations. The WHT is operated on the island of La Palma by the Isaac Newton Group in the Spanish Observatorio del Roque de los Muchachos 
of the Instituto de Astrofísica de Canarias. This publication makes use of data products from the Two Micron All Sky Survey, which is a joint 
project of the University of Massachusetts and the Infrared Processing and Analysis Center/California Institute of Technology, funded by the
National Aeronautics and Space Administration and the National Science Foundation. Funding for the SDSS and SDSS-II has been provided by the Alfred
P. Sloan Foundation, the Participating Institutions, the National Science Foundation, the U.S. Department of Energy, the National Aeronautics
and Space Administration, the Japanese Monbukagakusho, the Max Planck Society, and the Higher Education Funding Council for England. The SDSS Web
Site is http://www.sdss.org/. The SDSS is managed by the Astrophysical Research Consortium for the Participating Institutions. The Participating 
Institutions are the American Museum of Natural History, Astrophysical Institute Potsdam, University of Basel, University of Cambridge, Case 
Western Reserve University, University of Chicago, Drexel University, Fermilab, the Institute for Advanced Study, the Japan Participation Group, 
Johns Hopkins University, the Joint Institute for Nuclear Astrophysics, the Kavli Institute for Particle Astrophysics and Cosmology, the Korean 
Scientist Group, the Chinese Academy of Sciences (LAMOST), Los Alamos National Laboratory, the Max-Planck-Institute for Astronomy (MPIA), the 
Max-Planck-Institute for Astrophysics (MPA), New Mexico State University, Ohio State University, University of Pittsburgh, University of Portsmouth,
Princeton University, the United States Naval Observatory, and the University of Washington.
We thank the referee, Detlev Koester, for a prompt and helpful report.

\bsp

\label{lastpage}

\end{document}